    \documentstyle[proceedings,numreferences,epsfig]{crckapb} 

\newcommand{\bce}{\begin{center}}
\newcommand{\ece}{\end{center}}
\newcommand{\beq}{\begin{equation}}
\newcommand{\eeq}{\end{equation}}
\newcommand{\bea}{\begin{eqnarray}}

\newcommand{\eea}{\end{eqnarray}}

\newcommand{\ba}{\begin{array}}
\newcommand{\ea}{\end{array}}

\newcommand{\bDelta}{\mbox{\boldmath $\Delta$}}

\newcommand{\bkappa}{\mbox{\boldmath $\kappa$}}

\newcommand{\bb}{{\bf b}}
\newcommand{\br}{{\bf r}}

\newcommand{\bp}{{\bf p}}

\newcommand{\bs}{{\bf s}}

\def\lsim{\mathrel{\rlap{\lower4pt\hbox{\hskip1pt$\sim$}}
    \raise1pt\hbox{$<$}}}         
\def\gsim{\mathrel{\rlap{\lower4pt\hbox{\hskip1pt$\sim$}}
    \raise1pt\hbox{$>$}}}         
\def\beq{\begin{equation}}
\def\eeq{\end{equation}}
\def\bea{\begin{eqnarray}}
\def\eea{\end{eqnarray}}

\begin{opening}
\title{High Density QCD, Saturation and Diffractive DIS }
\author{
I.P. Ivanov}
\institute{Institute of Mathematics, Novosibirsk, Russia}
\author{{\underline{  N.N. Nikolaev}}$^{a,b)}$,W. Sch\"afer$^{a)}$, B.G. Zakharov$^{b)}$}
\institute{
$^{A)}$ Institut f. Kernphysik, Forschungszentrum J\"ulich, \\
D-52425 J\"ulich, Germany\\
$^{B)}$ L.D.Landau Institute for Theoretical Physics, \\
Chernogolovka, Russia}
\author{V.R. Zoller}
\institute{Institute for Theoretical and Experimental Physics, \\
Moscow, Russia
}
\end{opening}

\runningtitle{High Density QCD, Saturation and Diffractive DIS }

\begin{document}

\section*{Abstract}
We review a consistent description 
of the fusion and saturation of partons in the Lorentz-contracted
ultrarelativistic nuclei in terms of a nuclear attenuation of 
color dipole states of the photon and collective
Weizs\"acker-Williams (WW) gluon structure function of a nucleus.
Diffractive DIS provides a basis for the definition of the WW 
nuclear glue. The point that all observables for DIS off nuclei are 
uniquely calculable in terms of the nuclear WW glue amounts to
a new form of factorization in the saturation regime. 


\section{Introduction}

Within the QCD parton model the virtual photoabsorption
cross section is proportional to the density of partons in the
target and vice versa. 
When DIS is viewed in the laboratory frame, the 
hadronic properties of photons suggest \cite{NZfusion} a nuclear 
shadowing and depletion of the density of partons, when 
DIS is viewed in the Breit frame,
the Lorentz contraction of an ultrarelativistic  
nucleus entails a spatial 
overlap and fusion of partons at
$
x \lsim x_A={1/ R_A m_N} \sim 0.1\cdot A^{-1/3} \, .
$
This interpretation of nuclear opacity in terms of a fusion
and saturation of nuclear partons has been introduced in
1975 \cite{NZfusion} way before the QCD parton model. 
The pQCD link between nuclear opacity and
saturation has been considered in ref. \cite{NZ91} and 
by Mueller \cite{Mueller1},  the
pQCD discussion of fusion of nuclear gluons has been revived by
McLerran et al. \cite{McLerran}. 

Amplitudes of diffractive DIS are intimately related to the 
unintegrated glue of the target \cite{NZ92,NZsplit}. Because 
coherent diffractive DIS
in which the target nucleus does not break 
makes precisely 50 per cent of the total DIS events
for heavy nuclei at small $x$ \cite{NZZdiffr}, diffractive DIS
off nuclei offers a unique definition of the collective
WW glue of Lorentz-contracted ultarelativistic nuclei \cite{NSSdijet}.
The recent work has shown that all the observables for
nuclear DIS are uniquely calculable in terms of the NSS-defined
WW nuclear glue \cite{Saturation,JetJet}. In this overview 
presented at Diffraction'2002 by one of the authors (N.N.N) we 
summarize this new development.


\section{Quark and antiquark jets in 
DIS off free nucleons: single particle spectrum and jet-jet decorrelation}

In the  color dipole approach to DIS
\cite{NZ91,NZ92,NZZdiffr,NZ94,NZZlett,NZglue} the fundamental
quantity is the cross section for interaction of 
the $q\bar{q}$ dipole on a nucleon, 
\bea 
\sigma(r)= \alpha_S(r) \sigma_0\int d^2\bkappa
f(\bkappa )\left[1 -\exp(i\bkappa \br )\right]
\label{eq:2.1}
\eea 
where $f(\bkappa )$ is related to the unintegrated glue of
the target nucleon by 
\bea 
f(\bkappa ) = {4\pi \over
N_c\sigma_0}\cdot {1\over \kappa^4} \cdot {\partial G \over
\partial\log\kappa^2} \, .
\label{eq:2.2}
\eea
For DIS off a free nucleon, see figs. 1a-1d, 
$
\sigma_N = \int d^2\br dz |\Psi(z,\br)|^2 \sigma(\br)\, ,$ 
and the jet-jet inclusive cross section equals 
\bea
{d\sigma_N \over dz d^2\bp_+ d^2\bDelta} =
{\sigma_0\over 2}\cdot { \alpha_S(\bp^2) \over (2\pi)^2}
 f(\bDelta )
\left|\langle \gamma^*|\bp_+\rangle - 
\langle \gamma^*|\bp_+ -\bDelta \rangle\right|^2
\label{eq:2.9}
\eea
where $\bp_+$ is the transverse momentum of the quark,
$\bDelta= \bp_+ +\bp_-$ is the jet-jet decorrelation momentum,  $z_+= z$ and $
z_- = 1-z$
are the fractions of photon's lightcone momentum carried by the quark 
and antiquark, respectively, and the photon wave functions $\Psi(\br)$
and $\langle \bp |\gamma^*\rangle $ are found
in \cite{NZ91,NZ92}. The crucial point is that the jet-jet 
decorrelation is controlled \cite{Azimuth}
by the unintegrated gluon SF  $f(\bDelta )$.


\begin{figure}[!htb]
\begin{center}
\epsfig
{file=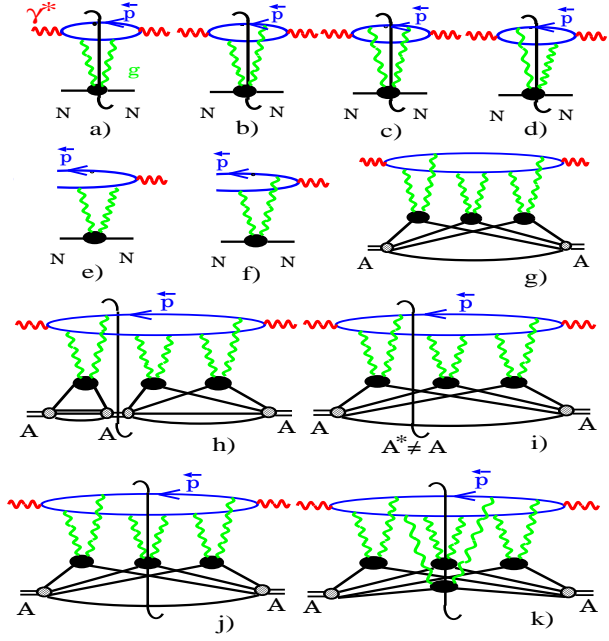, height=8.5cm, width = 8.0cm}
\end{center}
\caption{\it The pQCD diagrams for inclusive (a-d) and
diffractive
(e,f) DIS off protons and nuclei
(g-k). Diagrams (a-d) show the unitarity cuts with color
excitation of the target nucleon, (g) - a generic multiple
scattering diagram for Compton scattering off nucleus, (h) - the
unitarity cut for a coherent diffractive DIS, (i) - the unitarity
cut for quasielastic diffractive DIS with excitation of the
nucleus $A^*$, (j,k) - the unitarity cuts for truly inelastic DIS
with single and multiple color excitation of nucleons of the
nucleus. }
\end{figure}


\section{Non-Abelian propagation of color dipoles in nuclear me\-dium}
 
DIS at 
$x\lsim x_A$ is dominated by interactions 
of $q\bar{q}$ states of the photon. 
The unitarity cuts of the free-nucleon
forward Compton diagrams of figs. 1a-1d 
describe excitation from the color-neutral to color-octet $q\bar{q}$
pair.  The  unitarity cuts of the nuclear Compton scattering
amplitude of fig. 1g which correspond to diffractive and genuine 
inelastic DIS with color excitation of the nucleus are shown in fig. 1.
The elegant multichannel
description of the non-Abelian intranuclear evolution of color dipoles 
is found in \cite{JetJet}, here we cite only the principal
results.

Let $\bb_+$ and $\bb_-$ be the impact parameters of the quark
and antiquark, respectively, and $S_A(\bb_+,\bb_-)$  be the 
S-matrix for interaction of the $q\bar{q}$ pair with the nucleus.
We are interested in the 2-body (jet-jet) inclusive inelastic 
cross section when we sum 
over all color excitations of the target nucleus and all 
color states $c_{km}$ of the $q_k\bar{q}_m$ pair:
\bea
&&{d\sigma_{in} \over dz d^2\bp_+ d^2\bp_-} =
{1\over (2\pi)^4} \int d^2 \bb_+' d^2\bb_-' d^2\bb_+ d^2\bb_- \nonumber\\
&&\times \exp[-i\bp_+(\bb_+ -\bb_+')-i\bp_-(\bb_- -
\bb_-')]\Psi^*(\bb_+' -\bb_-')
\Psi(\bb_+ -\bb_-)\nonumber\\
&&\times \left\{\sum_{A^*} \sum_{km}  \langle
1;A|S_A^*(\bb_+',\bb_-')|A^*;c_{km}\rangle
\langle
c_{km};A^*|S_A(\bb_+,\bb_-)|A;1\rangle \right. \nonumber\\
&& - 
\left. \langle 1;A|S_A^*(\bb_+',\bb_-')
|A;1\rangle
\langle 1;A|S_A(\bb_+,\bb_-)|A;1\rangle \right\} \, ,
\label{eq:3.1} 
\eea
where the diffractive component
of the final state has been subtracted. 
According to \cite{Saturation,JetJet}, upon summing 
over nuclear final states $A^*$ and making use
of the
technique developed in \cite{NPZcharm,LPM},
the integrand of (\ref{eq:3.1}) can be represented as 
an $S$-matrix $S_{4A}(\bb_+,\bb_-,\bb_+',\bb_-')$ for 
the propagation of the two quark-antiquark pairs in the
overall singlet state. The detailed description 
of the matrix of 4-parton color dipole cross
section
$\sigma_4$ is found in \cite{JetJet}. The single-jet cross 
section evaluated at
the parton level equals
\cite{Saturation}
\bea {d \sigma_{in}\over d^2\bb d^2\bp dz }   
&=&  {1
\over (2\pi)^2}
 \int d^2\br' d^2\br
\exp[i\bp(\br'-\br)]\Psi^*(\br')\Psi(\br)\nonumber\\
&\times& \left\{\exp[-{1\over 2}\sigma(\br-\br') T(\bb)]-
\exp[-{1\over 2}[\sigma(\br)+\sigma(\br')]T(\bb)]\right\}
\, .
\label{eq:4.12} 
\eea


\section{The Pomeron-Splitting Mechanism for
Dif\-fractive Hard Dijets and Weizs\"acker-Wil\-l\-i\-ams glue of nuclei}
The two distinct diffractive dijet production         
QCD subprocesses are the classic Landau-Pomeranchuk-Feinberg-Glauber
beam-splitting  
\cite{Landau} (fig. 1e, fig. 2a)
and the Nikolaev-Zakharov pomeron-splitting \cite{NZ92,NZsplit}
 (fig. 1f, fig. 2b):
\begin{eqnarray}         
&&\Phi_0(z,\bp)=\, \int d^2\br          
\, e^{\displaystyle -i\bp \br} \, \sigma(\br)         
\,\Psi(z, \br) \nonumber\\
&&=\alpha_{S}(\bp^2)\sigma_{0}\left[         
\langle \bp|\gamma^*\rangle \int d^2\bkappa  f(\bkappa)         
- \int d^2\bkappa  \langle \bkappa|\gamma^*\rangle        
f(\bp-\bkappa)\right]         
\, ,         
\label{eq:5.2}         
\end{eqnarray}          
The amplitude for the former mechanism 
is $\propto \langle \bp|\gamma^*\rangle$
and the transverse momentum $\bp$ of jets comes from the intrinsic 
transverse momentum of $q,\bar{q}$ in the beam particle, in 
the latter jets receive          
a transverse momentum from gluons in the Pomeron, notice the 
convolution of the wave function with the unintegrated glue 
in the target proton.          
If the beam particle were a pion, then $\psi(z,\bp)$ would be much
steeper than $f(\bp)$, and the asymptotics of the convolution 
integral will be \cite{NSSdijet}         
\begin{equation}         
\int d^2\bkappa         
\psi_{\pi} (z,\bkappa)f(\bp-\bkappa) \approx f(\bp)\int d^2\bkappa         
\psi_{\pi} (z,\bkappa) = f(\bp) \, \phi_{\pi}(z)\, F_\pi  \, ,
\label{eq:5.6}        
\end{equation}         
and, furthermore, will probe the pion distribution amplitude $\phi_{\pi}(z)$.         

.
 
\begin{figure}[!htb]         
\begin{center}         
\epsfig
{file=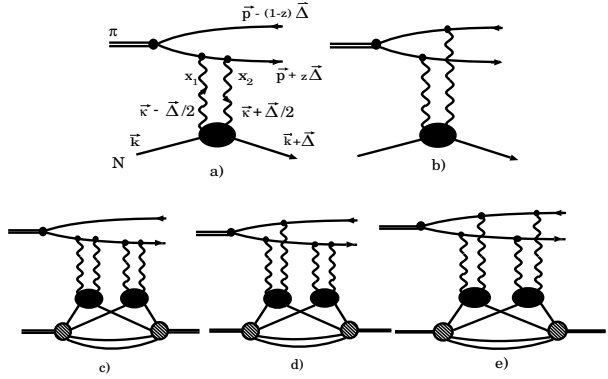, height = 8.0cm, width=5cm,angle=270}         
\end {center}         
\caption{\it Sample Feynman diagrams for diffractive dijet excitation  
in $\pi N$ collisions [diagrams 2a),2b)] and typical rescattering 
corrections to the nuclear coherent amplitude [diagrams 2c),2d),2e)].}         
\label{fig2}         
\end{figure}

Now we notice that in the color dipole representation
the nuclear amplitude is readily obtained \cite{NZ91} by substituting in         
eq.(\ref{eq:5.2}) 
\bea
\sigma(\br) &\to& \sigma_A(\br) =         
2 \, \int d^2\bb \{1-\exp[-\nu_{A}(\bb)]\}\\
\nu_{A}(\bb)&=& {1\over 2}\alpha_S(r)\sigma_0 T(\bb)= {1\over 2}\alpha_S(r)\sigma_0
\int dz n_A(\bb,z)
\label{eq:5.8}
\eea
Typical nuclear double scattering diagrams of figs.~2c-2e can         
be classified as  shadowing of the pion splitting (fig. 2c),         
shadowing of single Pomeron splitting (fig.~2d) and double Pomeron          
splitting (fig.~2e) contributions. In the $j$ Pomeron splitting         
$j$ exchanged Pomerons couple with one gluon to the quark and with one gluon          
to the antiquark of the dipole. That involves 
the $j$-fold convolution   
$
f^{(j)}(\bkappa )= \int \prod_{i=1}^j
d^2\bkappa _{i} f(\bkappa _{i}) \delta(\bkappa -\sum_{i=1}^j
\bkappa _i) \, .
$
Now we can invoke the NSS representation  \cite{NSSdijet,Saturation}
\bea
\exp\left[-\nu(\bb)\right] =\int d^2\bkappa \Phi(\nu_{A}(\bb), \bkappa)
\exp(i\bkappa\bs)
\label{eq:5.10}
\eea
where $
f^{(0)}(\bkappa)=\delta(\bkappa)$, $
\Phi(\nu_{A}(\bb), \bkappa) = \exp(-\nu_{A}(\bb))f^{(0)}(\bkappa)
+\phi_{WW}(\bb,\bkappa)$ and 
\bea
\phi_{WW}(\bb,\bkappa) =\exp\left[-\nu_{A}(\bb)\right]
\sum_{j=1}^{\infty} {1\over j!}\nu_{A}^{j}(\bb) 
f^{(j)}(\bkappa)
\label{eq:5.12}
\eea
can be identified with the unintegrated 
nuclear Weizs\"acker-Williams 
glue per unit area
in the impact parameter plane \cite{Saturation}. 


\section{Nuclear dilution and broadening of the unintegrated 
Weizs\"acker-Williams glue of nuclei}

The hard tail of WW glue per bound nucleon 
is calculable parameter free:
\bea
f_{WW}(\bb,\bkappa )= {\phi_{WW}(\bb,\bkappa)\over \nu_A(\bb)} 
= f(\bkappa )\left[1+ {2 C_A\pi^2\gamma^2\alpha_S(r)T(\bb)\over 
C_F N_c \bkappa^2}
G(\bkappa^2)\right] 
\label{eq:6.4}
\eea
In the hard regime the differential nuclear glue is not
shadowed, furthermore, because of the manifestly positive-valued
and model-independent nuclear higher twist correction it 
exhibits nuclear antishadowing property \cite{NSSdijet}.
The application of this formalism to the interpretation of
the E791 data on coherent diffraction of pions into dijets
\cite{Ashery} is found in \cite{NSSdijet,CEBAFtalk}.


\begin{figure}[!htb]
\begin{center}
   \epsfig
{file=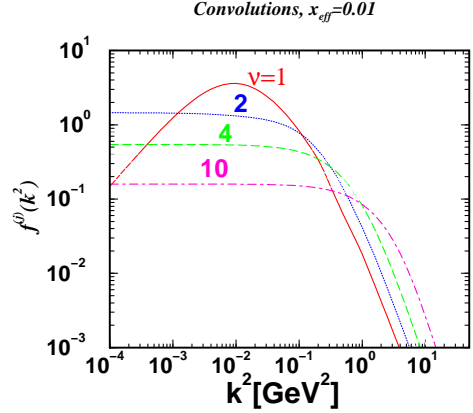,width=6cm}
\caption{\it The dilution for soft momenta and broadening for hard
momenta of the multiple convolutions {\protect $f^{(j)}(k)$} .}
\end{center}
\end{figure}

In the soft region one can argue that 
\bea
\phi_{WW}(\bkappa) \approx  {1\over \pi}  {Q_A^2 \over (\bkappa^2
+Q_{A}^2)^2}\, , 
\label{eq:6.15} \eea 
where the saturation scale $Q_A^2 =  \nu_A(\bb)  Q_0^2 \propto A^{1/3}\, . $
Notice a strong  nuclear dilution of soft WW glue, 
$
\phi_{WW}(\bkappa) \propto 1/Q_A^2 \propto
A^{-1/3}\, ,$
which must be contrasted to the $A$-dependence of hard WW glue 
(\ref{eq:6.4}), 
\beq
\phi_{WW}(\bkappa) = \nu_A(\bb) f_{WW}(\bkappa) \propto A^{1/3}\times
(1 + A^{1/3}\times(HT)+...)\, .
\label{eq:6.17}
\eeq

Take the platinum target, $A=192$. The numerical estimates based
on the parameterization \cite{INDiffGlue}  for $f(\bkappa)$ show that
for $(q\bar{q})$ color dipoles in the average DIS  on this target
$
\nu_{A} \approx 4
$ and 
$
Q_{3A}^2  \approx 0.8 {\rm GeV}^2\, .
$
For the $q\bar{q}g$ Fock states of the photon, which behave predominantly 
like  the dipole made of the two octet color charges, $Q_{8A}^2$ is larger 
by the factor $C_A/C_F = 9/4$, and for the the average DIS on the Pt 
target $ Q_{8A}^2 \approx 2.2$ GeV$^2$.


\section{Nuclear partons in the saturation regime}

On the one hand, making use of the NSS representation, the total
nuclear photoabsorption cross section 
(\ref{eq:5.8}) can be cast in the form 
\beq
\sigma_{A}  = \int d^2\bb\int dz \int {d^2\bp\over (2\pi)^2} \int
d^2\bkappa\phi_{WW}(\bkappa) \left|\langle \gamma^* |\bp\rangle -
\langle \gamma^* |\bp-\bkappa\rangle \right|^2 
\label{eq:8.1} 
\eeq
and its differential form defines the IS sea in a nucleus,
\bea {d\bar{q}_{IS} \over
d^2\bb d^2\bp} = {1\over 2}\cdot{Q^2 \over 4\pi^2 \alpha_{em}}
\cdot{d\sigma_A \over d^2\bb d^2\bp}\, . 
\label{eq:8.2} 
\eea 
Remarkably, in
terms of the WW nuclear glue, all intranuclear multiple-scattering
diagrams of fig.~3g sum up to precisely the same four diagrams
fig.~3a-3d as in DIS off free nucleons. 
On the other hand, making use of the NSS representation, after
some algebra one finds \cite{Saturation} 
\bea 
{d \sigma_{in}\over d^2\bb d^2\bp dz } &=&
{1 \over (2\pi)^2}\left\{ \int  d^2\bkappa
\phi_{WW}(\bkappa)\left|\langle \gamma^* |\bp\rangle -
\langle \gamma^* |\bp-\bkappa\rangle \right|^2 \right. \nonumber\\
&& - \left. \left|\int d^2\bkappa\phi_{WW}(\bkappa) (\langle
\gamma^* |\bp\rangle - \langle \gamma^*
|\bp-\bkappa\rangle )\right|^2\right\} \, ,
\label{eq:8.4} \\
{d \sigma_{D}\over d^2\bb d^2\bp dz }   &=&  {1 \over (2\pi)^2}
\left|\int d^2\bkappa\phi_{WW}(\bkappa) (\langle \gamma^*
|\bp\rangle - \langle \gamma^* |\bp-\bkappa\rangle)\right|^2 \, .
\label{eq:8.5} 
\eea 
Putting the inelastic and
diffractive components of the FS quark spectrum together, one
finds the FS parton density which exactly coincides with
the IS parton density (\ref{eq:8.2}) !

Consider first  $\bp^2 \lsim Q^2 \lsim Q_A^2$ when 
the nucleus is opaque for all color dipoles in the photon. 
In this regime the nuclear counterparts of
diagrams of figs. 1b,1d,1f  can be neglected and diffraction 
will be dominated by the the Landau-Pomeranchuk 
mechanism of fig.~1e,2a:
\bea 
\left. {d\bar{q}_{FS} \over d^2\bb d^2\bp}\right|_D  
\approx {Q^2 \over 8\pi^2 \alpha_{em}} \int
dz \left| \int d^2\bkappa\phi_{WW}(\bkappa) \right|^2
\left|\langle \gamma^* |\bp\rangle\right|^2 \approx {N_c\over 4\pi^4}\, . 
\label{eq:8.6} 
\eea 
Remarkably, diffractive DIS measures the momentum distribution 
in the $q\bar{q}$ Fock state of the photon.
 In
contrast to diffraction off free nucleons
\cite{NZ92,NZsplit,GNZcharm}, diffraction off opaque nuclei is
dominated by the anti-collinear splitting of hard gluons into soft
sea quarks, $\bkappa^2 \gg \bp^2$. Precisely for this reason one
finds the saturated FS quark density, because the nuclear dilution
of the WW glue is compensated for by the expanding plateau.

The related analysis of the FS quark density for truly inelastic
DIS in the  same domain of $\bp^2 \lsim Q^2 \lsim Q_A^2$ gives
\bea 
&&\left.{d\bar{q}_{FS} \over d^2\bb d^2\bp}\right|_{in} =
{1\over 2}\cdot{Q^2 \over 4\pi^2 \alpha_{em}} \cdot\int dz \int
d^2\bkappa \phi_{WW}(\bkappa)
\left|\langle \gamma^* |\bp-\bkappa\rangle \right|^2 \nonumber\\
&&\approx
{Q^2 \over 8\pi^2 \alpha_{em}}\phi_{WW}(0)
\int^{Q^2} d^2\bkappa \int dz \left|\langle \gamma^* |\bkappa\rangle \right|^2 
\nonumber\\
&&= {N_c \over 4\pi^4}\cdot {Q^2 \over Q_A^2}\cdot\theta(Q_A^2-\bp^2)
\label{eq:8.7}
\eea
which, as a functional of the photon wave
function and nuclear WW gluon distribution, is completely
different from the free-nucleon version of (\ref{eq:8.1}). 

\begin{figure}[!htb]
   \centering
   \epsfig
{file=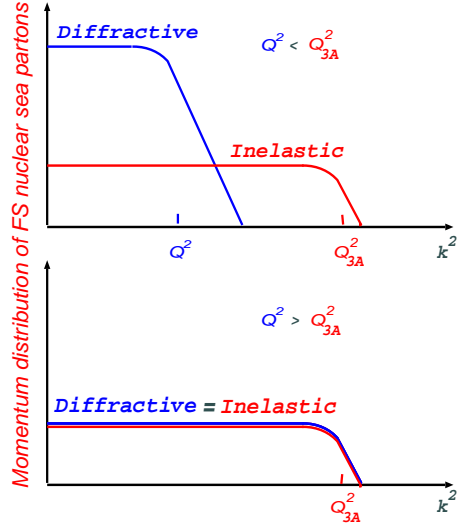,width=6cm,height=7cm}
\caption{\it The two-plateau structure of the 
momentum distribution of FS quarks: (i)
$Q^2 \lsim Q_{3A}^2$ : The inelastic  plateau is much broader than
the diffractive one. (ii)
$Q^2 \gsim Q_{3A}^2$ : inelastic  plateau is identical to diffractive one.
} 
\end{figure}

Now notice, that  in the opacity regime the diffractive FS parton
density coincides with the contribution $\propto |\langle
\gamma^*|\bp\rangle|^2$ to the IS sea parton density from the
spectator diagram 1a, whereas the FS parton density for truly
inelastic DIS coincides with the contribution to IS sea partons
from the diagram of fig.~1c. The contribution from the crossing
diagrams 1b,d is negligibly small.
Nuclear broadening
and unusually strong $Q^2$ dependence of the FS/IS parton density from
truly inelastic DIS, demonstrate clearly a distinction between diffractive
and inelastic DIS, see fig. 4. 

In the case of soft quarks, $\bp^2 \lsim Q_A^2$, in hard photons, 
$Q^2 \gsim Q_{A}^2$, the result (\ref{eq:8.6}) for diffractive DIS is retained,
whereas in the numerator of the result (\ref{eq:8.7}) for truly inelastic
DIS one must substitute $Q^2 \to Q_{A}^2$, so that in this case
$d\bar{q}_{FS}|_{D} \approx d\bar{q}_{FS}|_{in}$ and $d\bar{q}_{IS} \approx
2d\bar{q}_{FS}|_{D}$, see fig. 4. The evolution
of soft nuclear sea, $\bp^2 \lsim Q_{A}^2$, is entirely driven by
an anti-collinear splitting of the NSS-defined WW nuclear glue into the sea
partons.

The early discussion of the FS quark density in the saturation
regime is due to Mueller \cite{Mueller}. Mueller focused on $Q^2
\gg Q_A^2$ and discussed  neither a distinction between
diffractive and truly inelastic DIS nor a $Q^2$ dependence and
broadening (\ref{eq:8.7}) for truly inelastic DIS at $Q^2 \lsim
Q_A^2$.


\section{Signals of saturation in exclusive diffractive DIS}

The flat $\bp^2$ distribution of forward $q,\bar{q}$ jets 
in truly inelastic DIS in the saturation regime must be contrasted 
to the $\propto G(\bp^2)/\bp^2$ spectrum for the free nucleon
target. In the diffractive DIS the saturation gain is much more
dramatic: flat $\bp^2$ distribution of forward $q,\bar{q}$ jets 
in diffractive DIS in the saturation regime must be contrasted 
to the $\propto 1/(\bp^2)^2$ spectrum for the free nucleon
target \cite{NZ92,NZsplit,GNZcharm}. In the general case one must
compare the saturation scale $Q_A$ to the relevant hard scale $\bar{Q}^2$
for the specific diffractive process.  
For instance, in exclusive diffractive DIS, i.e., the vector meson
production, $\bar{Q}^2 \approx (Q^2+m_V^2)/4$
and the transverse cross section has been predicted to behave
as \cite{NNZscan}
\beq
\sigma_T \propto G^2(x,\bar{Q}^2)(\bar{Q}^2)^{-4}
\label{eq:9.2}
\eeq
At $\bar{Q}^2 > Q_A^2$ the same would hold for nuclei too, 
but in the opposite case of $\bar{Q}^2 < Q_A^2$ the 
$\bar{Q}^2$-dependence is predicted to change to 
\beq
\sigma_T \propto G^2(x,\bar{Q}^2)(Q_A^2)^{-2}
(\bar{Q}^2)^{-2}
\label{eq:9.3}
\eeq


\section{Jet-jet decorrelation in DIS off nuclear targets}

The derivation of the jet-jet inclusive cross section requires
a full fledged non-Abelian 
multichannel calculation of $S_{4A}(\bb_+,\bb_-,\bb_+',\bb_-')$.
The principal result for the hard, $\bp_{\pm}^2 \gg Q_A^2$,
 jet-jet cross section is \cite{JetJet}
\bea
{d\sigma_{in} \over d^2\bb dz d^2\bp_+ d^2\bDelta}= T(\bb)
\int_0^1 d \beta \int d^2\bkappa \Phi(2\beta \lambda_c  \nu_A(\bb),\bDelta - \bkappa)
{d\sigma_{N} \over dz d^2\bp_+ d^2\bkappa }\, .
\label{eq:10.17} 
\eea  
It has a probabilistic form of a convolution of the differential cross 
section on a free nucleon target with the unintegrated nuclear WW 
glue $\Phi(2\beta \lambda_c  \nu_A(\bb),\bkappa)$, $\beta$ has a meaning 
of the fraction of the nuclear thickness which the $(q\bar{q})$
pair propagates in the color-octet state and the nontrivial color factor
$\lambda_c = (N_c^2 - 2)/(N_c^2 -1)$
is a manifestation of the non-Abelian intranuclear propagation of color dipoles.

The azimuthal decorrelation of hard jets is quantified by 
\beq
\langle \bDelta_{\perp}^2 \rangle \approx  {Q_A^2 \over 2} 
\left( {Q_A^2 +\bp_+^2 \over \bp_{+}^2} \log{Q_A^2 +\bp_+^2 \over  Q_A^2} -1\right)
\label{eq:10.21}
\eeq
and the differential out-of-plane momentum distribution 
\beq
{d\sigma \over d\Delta_{\perp}} \approx 
{1\over 2} {Q_A^2 \over (Q_A^2 +\Delta_{\perp^2})^{3/2}}
\label{eq:10.22}
\eeq
which shows that the probability to observe the away jet at $\Delta_{\perp}
\sim 0$ disappears $\propto 1/Q_A \propto 1/\sqrt{\nu_A}$.
For instance, from peripheral DIS to
central DIS on a heavy nucleus like $Pt$, $\nu_{8A}$ rises 
form $\nu_{8A}=1$ to $\nu_{8A}\sim 13$, so that according 
to (\ref{eq:10.22}) a probability to find the away jet
decreases by the large factor $\sim 3.5$ from peripheral
to central DIS off $Pt$ target.  

In the generic case the closed form for the jet-jet inclusive 
cross section is found \cite{JetJet}  only for large $N_c$:
\bea
{d\sigma_{in} \over d^2\bb dz d\bp_{-} d\bDelta} &=&
{1\over 2(2\pi)^2} \alpha_S \sigma_0 T(\bb)\int_0^1 d \beta 
\int d^2\bkappa_1 d^2\bkappa_2 d^2\bkappa_3 d^2\bkappa 
f(\bkappa) \nonumber\\
&\times &\Phi((1-\beta)\nu_A(\bb),\bkappa_1) 
\Phi((1-\beta)\nu_A(\bb),\bkappa_2) \nonumber\\
&\times &
\Phi(\beta\nu_A(\bb),\bkappa_3) 
\Phi(\beta\nu_A(\bb),\bDelta -\bkappa_3 -\bkappa)\nonumber\\
&\times &\left\{\Psi(-\bp_{-} +\bkappa_2 +\bkappa_3)-
\Psi(-\bp_{-} +\bkappa_2 +\bkappa_3+\bkappa)\right\}^*\nonumber\\
&\times &
\left\{\Psi(-\bp_{-} +\bkappa_1 +\bkappa_3)-
\Psi(-\bp_{-} +\bkappa_1 +\bkappa_3+\bkappa)\right\}
\label{eq:11.5}
\eea
We emphasize that it is uniquely calculable in terms of the 
NSS-defined WW glue of the nucleus.

One important implication of (\ref{eq:11.5}) is that for minijets
with the jet momentum $|\bp_-|,|\bDelta| \lsim Q_A$, the 
the minijet-minijet inclusive cross 
section depends on neither the minijet nor decorrelation momentum, 
i.e., we predict a complete disappearance of the azimuthal decorrelation 
of jets with the transverse momentum below the saturation scale.

Our experience with application
of color dipole formalism to hard hadron-nucleus interactions
\cite{NPZcharm,LPM} suggests that our analysis can be readily
generalized to mid-rapidity jets. For this reason, we
expect similar strong decorrelation of mid-rapidity jets
in hadron-nucleus and nucleus-nucleus collisions. To this end,
recently the STAR collaboration reported a disappearance of
back-to-back high $p_{\perp}$ hadron correlation in central
gold-gold collisions at RHIC \cite{STARRHIC}. Nuclear
enhancement of the azimuthal decorrelation of the trigger 
and away jets may 
contribute substantially to the STAR effect. 

\section{Summary and conclusions}

We reviewed a theory of DIS off nuclear targets based on the
consistent treatment of propagation of color dipoles in nuclear
medium. What is viewed as attenuation in the laboratory frame can
be interpreted as a fusion of partons from different nucleons of
the ultrarelativistic nucleus. Diffractive attenuation of color
single $q\bar{q}$ states gives a consistent definition of the 
WW unintegrated gluon structure function of the nucleus
\cite{NSSdijet,Saturation}, all 
other nuclear DIS observables - sea quark structure function
and its decomposition into equally important genuine inelastic 
and diffractive components, exclusive diffraction off nuclei,
the jet-jet inclusive cross section,
- are uniquely calculable in terms of the NSS-defined 
nuclear WW glue. This
property can be considered as a new factorization which connects
DIS in the regimes of low and high density of partons. 
The anti-collinear splitting of WW nuclear glue is a clearcut
evidence for inapplicability of the DGLAP evolution to nuclear
structure functions unless $Q^2 \gg Q_{8A}^2$.  

N.N.N. is grateful to L. Jenkovszky for the invitation to
Diffraction'2002.
This work has been partly supported by the INTAS grants 97-30494
\& 00-00366 and the DFG grant 436RUS17/89/02.



\end{document}